# Unidirectional single-photon emission from germanium-vacancy zero-phonon lines: Deterministic emitter-waveguide interfacing at plasmonic hot spots

Hamidreza Siampour[1*], Ou Wang[2], Vladimir A. Zenin[1], Sergejs Boroviks[1], Petr Siyushev[2], Yuanqing Yang[1], Valery A. Davydov[3], Liudmila F. Kulikova[3], Viatcheslav N. Agafonov[4], Alexander Kubanek[2], N. Asger Mortensen[1,5], Fedor Jelezko[2], and Sergey I. Bozhevolnyi[1,5]

[1] Centre for Nano Optics, University of Southern Denmark, Campusvej 55, Odense M, DK-5230, Denmark
[2] Institute for Quantum Optics, Ulm University, Albert-Einstein-Allee 11, D-89081 Ulm, Germany
[3] L.F. Vereshchagin Institute for High Pressure Physics, Russian Academy of Sciences, Moscow, 142190, Russia
[4] GREMAN, UMR CNRS CEA 6157, Université de Tours, 37200 Tours, France
[5] Danish Institute for Advanced Study, University of Southern Denmark, Campusvej 55, Odense M, DK-5230, Denmark

[*]hasa@mci.sdu.dk

**ABSTRACT-** Striving for nanometer-sized solid-state single-photon sources, we investigate atom-like quantum emitters based on single germanium vacancy (GeV) centers isolated in crystalline nanodiamonds (NDs). Cryogenic characterization indicated symmetry-protected and bright ($> 10^6$ counts/s with off-resonance excitation) zero-phonon optical transitions with up to 6-fold enhancement in energy splitting of their ground states as compared to that found for GeV centers in bulk diamonds (i.e., up to 870 GHz in highly strained NDs vs 150 GHz in bulk). Utilizing lithographic alignment techniques, we demonstrate an integrated nanophotonic platform for deterministic interfacing plasmonic waveguides with isolated GeV centers in NDs that enables 10-fold enhancement of single-photon decay rates along with the emission direction control by judiciously designing and positioning a Bragg reflector. This approach allows one to realize the unidirectional emission from single-photon dipolar sources introducing a novel method that is alternative to the propagation-direction-dependent techniques based on chiral interactions or topological protection. The developed plasmon-based nanophotonic platform opens thereby new perspectives for quantum nanophotonics in general and for realizing entanglement between single photons and spin qubits, in particular.

**INTRODUCTION**

Efficient interfaces between single atoms and single photons are essential ingredients for building quantum optical networks, where atomic nodes (quantum emitters) are linking together via flying photons (qubits)[1,2]. The key challenge here is to engineer atom-photon interactions in order to have control over individual quantum emitters on a large scale. Combining isolated atoms with nanophotonic systems is a powerful approach to strongly enhance the atom-photon interaction due to a large cooperativity associated with nanoscale photonic devices, although trapping atoms in tightly focused laser beams (optical tweezers) imposes serious technical challenges[3-5]. Alternatively, an all solid-state approach has been developed, in which naturally trapped "atoms" (e.g., quantum dots) are coupled, created and multiplexed on a single chip[6]. Despite the progress made for deterministic positioning of quantum dots on a single chip[7-10], some challenges remain due to a relatively short coherence time available with quantum dots (in nanosecond range)[11], resulting in the quantum information being typically lost before reaching distant quantum nodes. Further search for configurations insuring long coherence times in atomic systems and allowing for scalable implementation in solid-state systems led to the exploration of diamond crystals containing artificial "atoms" (so-called color centers)[12]. Starting with a nitrogen-vacancy center (i.e., substitutional nitrogen-atom impurity next to a diamond lattice vacant site), remarkable coherence time (in millisecond range)[13] has been achieved, making it an ideal emitter for spin physics and metrology[14]. However, a lack of symmetry in nitrogen-vacancy molecule structure limits severely the coherent part of the emission, with the emission to a zero-phonon line (ZPL) being only 4%, and makes the frequency of optical transitions being very sensitive to the environment. Replacing nitrogen with larger atoms of group IV in the periodic table (e.g., with a silicon atom that is a ~1.5 times larger in size than a carbon atom) enabled to circumvent the issues associated with symmetry arguments[15-20] (see Figure 1a,b). This opened a way toward demonstrations of indistinguishable solid-state quantum emitters (without the need for electric field tuning) with spectral stability and large ZPLs[21]. However, the efficiency of photon out-coupling from diamond color centers in bulk crystals is limited (due to a high refractive index of diamond)[12,16,17,22-24].

Progressing towards bright and efficient nanometer-sized solid-state single-photon sources, here we report on the investigation and control of atom-like quantum emitters based on germanium vacancy (GeV) centers isolated in crystalline nanodiamonds (NDs). Cryogenic characterization indicated symmetry-protected and bright (> $10^6$ counts/s with off-resonance excitation) zero-phonon optical transitions featuring remarkable energy splitting in their ground states, up to 870 GHz that is ~6 times larger than that in bulk diamonds. The large energy split in the ground state implies a potentially longer spin coherence due to the suppressed phonon-mediated transitions between the lower and upper branches[17,25]. This opens new perspectives for deterministic interfacing of isolated atoms with photons along with merging quantum emitters with highly confined surface plasmons in metal-based nanostructures[5,26,27]. Utilizing lithographic alignment techniques[27-30], we demonstrate an integrated nanophotonic platform for deterministic interfacing plasmonic waveguides with isolated GeV centers in NDs that enables 10-fold enhancement of single-photon decay rates along with the emission direction control by judiciously designing and positioning a Bragg reflector. This approach allows one to realize the unidirectional emission from single-photon dipolar sources introducing a novel method that is alternative to the propagation-direction-dependent techniques based on chiral interactions[31,32] or topological protection[33]. The developed plasmon-based nanophotonic platform opens thereby new perspectives for quantum nanophotonics, particularly for realizing entanglement between single photons and spin qubits[34].

**RESULTS AND DISCUSSION**

Imitating the natural formation of diamonds underneath the Earth, diamond crystals were grown at the scale of nanometer, under a high-pressure high-temperature (HPHT) condition, and Ge defect atoms were added during the growth in a hydrocarbon metal catalyst-free system based on homogeneous mixtures of naphthalene $C_{10}H_8$ with tetraphenylgermanium $C_{24}H_{20}Ge$ (see details in Supplementary, Section 1). Cryogenic characterization shows symmetry-protected optical transitions for the synthesized GeV centers in NDs (see Figure 1c), following the reported trend for bulk diamonds[16-19] (see Figure 1b). Furthermore, ZPLs indicate a large splitting in the ground state (up to 870 GHz), which is ~17 times larger than silicon-vacancy (SiV)[19], and ~6 times larger than GeV in bulk[16,35,36], becoming close to tin-vacancy (SnV) with 850 GHz[17]. Power dependency measurements at low temperature (shown in Figure 1d) exhibit ultrabright single photon count rates ($> 2\times10^6$ counts/s at saturation) at ZPLs (i.e., excluding phonon sideband) with clean single photon emission (strong antibunching dip of $g^2(0)=0.06$, where $g^2(0)$ is the autocorrelation function at zero-delay time). Deterministic loading of dielectric nanostructures on colloidal gold crystals enabled to trap a pre-selected ND at a plasmonic hot spot for realizing a sustainable nanoplasmonic platform for integrated quantum photonics. This introduces a powerful approach, alternative to the current deterministic techniques (atomic-force microscope (AFM) manipulation[24,37], site-controlled and strain-induced excitation[8], in-situ cathodoluminescence lithography[9], and optical trapping[3,5]) to improve the level of control on coupling and, at the same time, to facilitate scalable fabrication (see Figure 2).

In the experiment, colloidal gold crystals were grown on a silicon substrate using a thermolysis synthesis technique[38] (see details in Supplementary, Section 2). Lithographic alignment markers were made on top of (and nearby) the gold crystals, and NDs were deposited on the substrate afterwards. The sample was then loaded on the cold-finger of a continuous flow helium cryostat, which was cooled to 4.7 K for confocal microscopy measurements (see Methods, Section 2). Fluorescence image was taken from a crystalline gold flake on which NDs containing single GeV centers were deposited (see Figure 4S in Supplementary for the corresponding confocal image). A single GeV emitter was selected based on fluorescence spectrum measurements, and its location was determined with respect to the markers. Utilizing lithographic alignment techniques[27-30], we fabricated a waveguide-mirror configuration with reflecting Bragg gratings (RBG), and the pre-selected ND being embedded in the waveguide and placed at the specific constructive interference point (at the distance of the second constructive point of the RBG mirror, i.e., $d = 3\lambda_n/4$, where $\lambda_n$ is the wavelength of the plasmonic mode[39]) as shown in Figure 3a-c. Excited with the green laser ($\lambda$ = 532 nm), the fluorescence light was collected from the excitation point (GeV-ND) and away of it within several micometers using a galvanometric scanning mirror. The result is illustrated in Figure 3d, showing two spots, one from the excited GeV emitter, and the other from the grating output at the end (outcoupler). This indicated coupling of GeV emission to the waveguide mode (so-called dielectric-loaded surface plasmon polariton mode[40,41]) of the unidirectional SPP device. Fluorescence spectrum taken at the output grating was compared with the spectrum from the center (i.e., from part of the emission that is not coupled to the plasmon mode), indicating a narrower linewidth at the end (see Figure 3f). This can be explained with polarization properties of the propagating plasmon mode, for which out-of-plane component is dominant (see Supplementary, Figure 16S). Therefore, only compatible part of the emission (that is polarized either out-of-plane or along waveguide axis) is allowed to propagate along the plasmonic waveguide while the remaining power contributed to the far-field radiation collected at the center. Polarization selectivity of the unidirectional coupler was simulated by modeling the GeV emitter by an electric dipole oriented along different directions to excite the propagating mode in the waveguide (see Methods section and Figure 16S in Supplementary). The result for the out-of-plane dipole component is presented in Figure 3e, indicating a good agreement with the experimental result shown in Figure 3d. The direction of propagating single photons emitted from the GeV center (and in general all group IV color centers[16,17,19,24]) can be controlled by using this way, an alternative approach to other propagation-direction-dependent techniques based on chiral interactions[31,32] or topological protection[33]. Lifetime measurements before and after structure fabrication (Figure 3g,h) indicate a 10-fold enhancement in the total decay rate of the GeV emitter, a remarkable record due to the interaction with the unidirectional SPP coupler (see Figure 5S-11S in Supplementary for further details and additional experiments).

In order to verify the properties of the unidirectional SPP coupler, we performed scanning near-field optical microscope (SNOM) measurements with a titanium sapphire laser source (wavelength 775-1000 nm). Because our SNOM is operating in a transmission configuration with sample illuminated from below, a grating was fabricated in the gold layer for efficient excitation of plasmonic mode (see schematic in Figure 4a and details in Methods section). An SEM image of the fabricated device and an AFM image of the excitation point are shown in Figure 4b and Figure 4c, respectively. The RBG mirror period was modified to produce strong back-reflection at the central operating wavelength (Figure 4d, $\lambda = 850$ nm), while no reflection was observed outside the transverse magnetic (TM) bandgap (Figure 4e, $\lambda = 1000$ nm). Mode parameters (effective mode index, propagation length, and reflection coefficient) were extracted from near-field maps, obtained for different laser wavelengths, using a simple fitting procedure[42-44] (see Supplementary, Figures 13S and 14S). The resulting SNOM measurements indicate ~80% reflectance from the RBG mirror (Figure 4g).

## CONCLUSIONS

Single GeV centers in diamond nanocrystals have been investigated at low temperature, indicating symmetry-protected and ultrabright ($> 2\times10^6$ counts/s at saturation) zero-phonon optical emission. A large energy split of 870 GHz in the ground state has been determined for the GeV centers in highly strained ND samples, that is ~6 times larger than that in bulk diamonds. An integrated crystalline gold-based nanophotonic platform has been realized, on which GeV centers in NDs interface with highly confined SPP modes in a unidirectional manner, resulting in a remarkable 10-fold Purcell enhancement. The unidirectional interaction enables efficient guiding control of the emitted single photons, revealing the potential of our approach for realizing entanglement between an optical photon and a spin qubit.

## METHODS

**Device fabrication.** Direct e-beam writing was performed using a negative tone electron-beam resist comprised of 6% solution of hydrogen silsesquioxane (HSQ) diluted in methylisobutylketone (MIBK) solvent (Dow Corning XR-1541−006). Using a standard spin-on coating equipment, HSQ was deposited on the substrate with the speed of 1200 rpm (1 min), and subsequently the solvent was boiled off during a hotplate bake process (170ºC, 2 min). This resulted in a 180 nm film on gold crystal flakes. Using a single exposure tool with 30 KeV beam energy and area doses from 400 to 700 µC/cm$^2$, the HSQ film was patterned, with capability to define features as small as 6 nm, and nanoridge waveguides were defined and accurately positioned onto preselected NDs, whose locations were determined with respect to the specifically designed and prefabricated alignment markers. Then, the HSQ film was developed in tetramethylammonium hydroxide (Sigma-Aldrich, 25 wt. % solution in water), a standard aqueous base developer (see further details in Figure 5S in Supplementary).

**Cryogenic measurements.** The sample was mounted on the cold-finger of a continuous flow helium cryostat, which was cooled to 4.7 K for imaging with a home-built confocal microscope (see details in Supplementary, Figure 3S). Experimental control was provided by the Qudi software suite[45]. The GeV centers were off resonantly excited by linearly polarized 532 nm green laser to map the fluorescence of GeV ZPL. Band pass filter (599/13 nm) was placed in front of the avalanche photodiode (APD). Spectra were measured after a 560 nm long pass filter (Supplementary, Figure 3S).

**Scanning near-field microscopy (SNOM).** Near-field investigation was performed using commercial AFM-based scattering-type SNOM (NeaSpec). In this setup, used in transmission mode, the sample was illuminated normally from below (focused by a bottom parabolic mirror with NA ~ 0.3) using a tunable CW Ti:Sapphire laser (wavelength 775-1000 nm, Spectra Physics). The illumination was coupled into the waveguiding SPP mode by means of a grating in the gold layer. The sample was scanned in a noncontact mode by a commercial Pt-coated AFM tip (Arrow-NCPt from NanoWorld), which acts as a scatterer of the near-field. The tapping amplitude and frequency were ~60 nm and $\Omega$ ~ 250 kHz, correspondingly. The tip's scattered light was collected by a top parabolic mirror (NA ~ 0.7) and forwarded towards photodetector. In order to resolve both amplitude and phase, a Mach-Zehnder interferometer was integrated in our setup, in which the optical path difference was modulated by an oscillating mirror in the reference arm (frequency $f$ ~ 300 Hz). The detected signal was then demodulated using pseudo-heterodyne detection method[46] (see further details in Supplementary, Figure 12S-14S).

**Numerical simulations.** Finite-difference time-domain (FDTD) simulations were performed using a commercial software (FDTD Solutions v8.11.318, Lumerical). Perfectly matched layers (PMLs) were applied to enclose the computational domain of 8.5 µm × 3 µm × 2 µm in all calculations. The finest mesh grid size of 5 nm is used in the simulations. The optical constants of the gold were taken from the experimental data reported by Johnson and Christy[47]. To mimic the experimental conditions, we utilized an electric dipole oriented along different directions to excite the propagating mode in the waveguide. The distance between the dipole and the RBG mirror was set as $d = 3\lambda_n/4$ for an efficient unidirectional excitation while the height of the dipole was set to 30 nm. To compare with the experimental observations, we implemented far-field projection to simulate the far-field images observed by a microscope with 0.95 numerical aperture[48] (see Supplementary, Figure 16S). Furthermore, we simulated the characteristics of the SPP modes at cryogenic temperatures using finite element method (Supplementary, Section 14).

## ACKNOWLEDGMENTS

This work was supported by the European Research Council (ERC), Advanced Grant 341054 (PLAQNAP). FJ acknowledges support of the DFG, BMBF, VW Stiftung and EU (ERC, DIADEMS). VAZ, SB, and NAM acknowledges support of VILLUM FONDEN (Grant 16498). VAD and LFK thank the Russian Foundation for Basic Research (Grant No. 18-03-00936) for financial support. AK acknowledges support of the DFG, the Carl-Zeiss Foundation, IQST, the Wissenschater-Ruckkehrprogramm GSO/GZS.

## CONTRIBUTIONS

HS, SIB and FJ conceived the experiment. HS performed the device fabrication, measurements, and computations. OW and PS contributed to the cryogenic measurements. VAZ performed SNOM measurements. SB synthesized the colloidal gold crystals and NAM supervised his experiment. YY contributed to the FDTD simulations. VAD, LFK and VNA supplied the nanodiamond samples. HS wrote the manuscript with contributions from all authors. All authors discussed the results and commented on the manuscript.

## COMPETING INTERESTS

The authors declare no competing interests.

## FIGURE LEGENDS

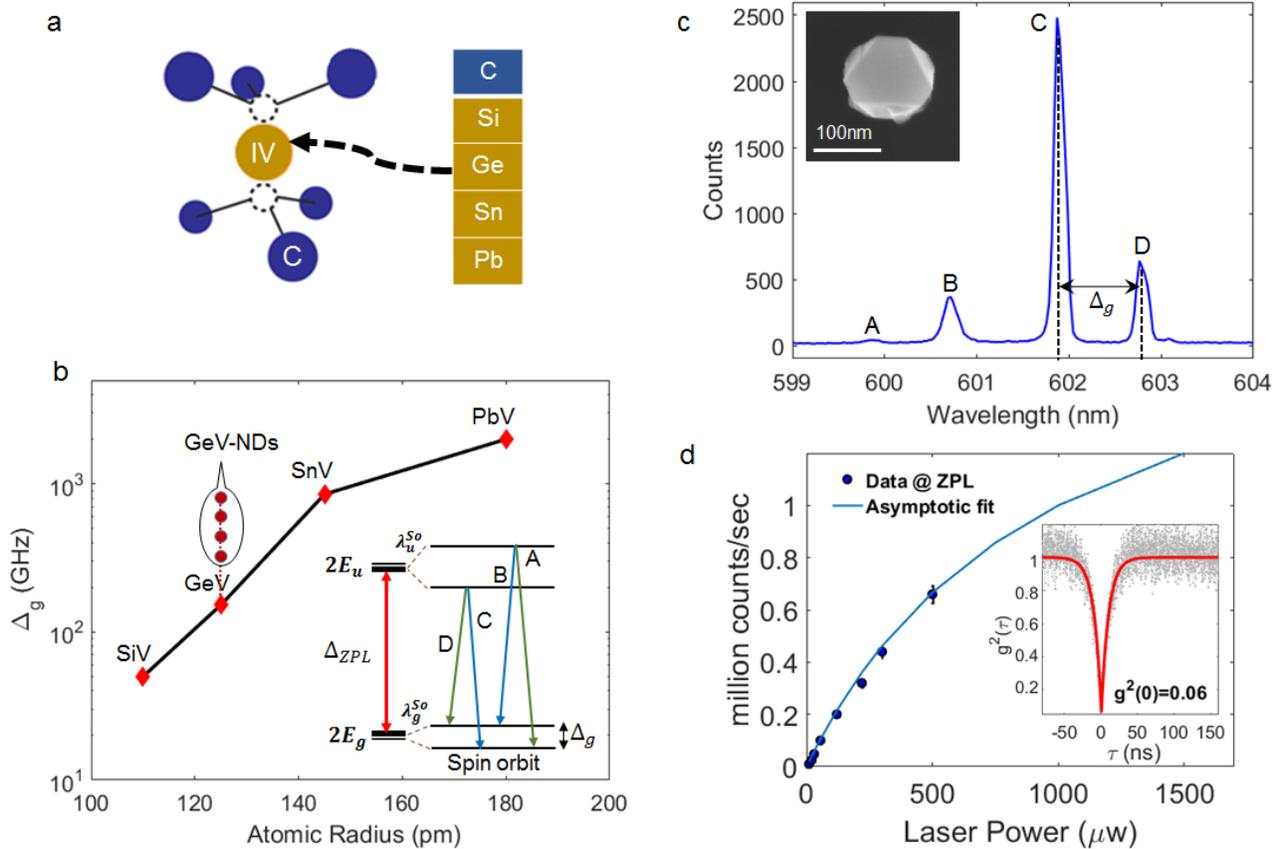

**Figure 1** Color centers in diamond crystals with structural symmetries and bright ZPLs. (**a**) The defect atom in group IV (silicon, germanium, tin, or lead) is placed between two diamond lattice vacant sites, resulting in a split-vacancy color center (SiV, GeV, SnV, or PbV) with spectral stability as a result of its inversion symmetry. (**b**) Energy split in the ground state ($\Delta_g$) for group IV color centers in diamond. The larger atom exhibits larger splitting and potentially a longer spin coherence time. The values of $\Delta_g$ for color centers in bulk diamond, including SiV[19], GeV[16], SnV[17] and PbV[18], were adapted from the experimental results reported for the corresponding vacancies. Having GeV centers in NDs results in even larger energy split in the ground state due to the strain conditions in nanocrystals. Inset shows zero-phonon optical transition lines for a typical group IV color center. Internal transitions of B and C are parallel in polarization and orthogonal to the external transitions of A and D, resulting in an emitter with two orthogonal dipoles[19,24]. (**c**, **d**) Characterization of a single GeV center in an HPHT ND. Spectrum taken at a cryogenic temperature exhibits four-line fine structure at around 602 nm similar to bulk crystals, but with ~6 times larger ground state splitting of $\Delta_g$=870 GHz (**c**). Inset shows SEM image of the ND. The temperature of the ND was calculated to be ~40 K using Boltzmann statistics[49], which is higher than the temperature of the cryostat cold finger due to the limited thermal conductivity of the substrate (silicon). (**d**) Power dependency measurements indicates ultrabright ZPLs (exceeding 1 million counts per second at 1mW). Saturation curve was fitted to an asymptotic function ($I = I_\infty \cdot P/(P + P_\infty)$, where $I_\infty$ and $P_\infty$ are saturated intensity and saturated power, respectively), indicating a 15-fold enhancement in the brightness compared to those reported for bulk diamonds[36]. Inset shows strong antibunching dip in autocorrelation measurement ($g^2(0)$=0.06), which implies single photon emission. The red line represents a single exponential fit[16].

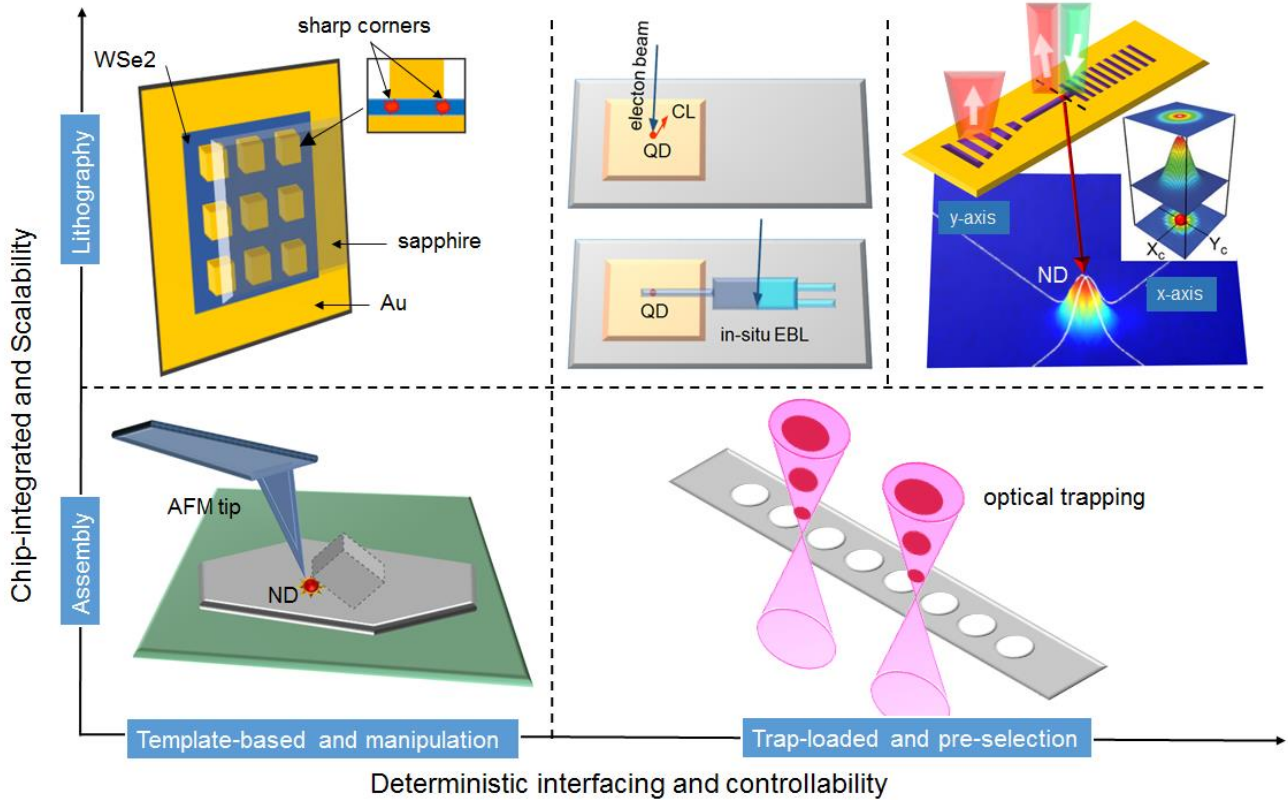

**Figure 2** Advancements in nanotechnologies for deterministic integration of single quantum emitters at nanoscale photonic hot spots. Nanoscale optical positioning techniques[28] were utilized to accurately find the location of a single emitter in a ND. Using lithographic alinement techniques, dielectric nanoridges were fabricated atop gold crystals, so that the pre-selected ND is embedded in the nanoridge and positioned in the second constructive interference fringe of the RBG mirror (hot spot). This technique is compared with other deterministic technologies (e.g., AFM manipulation[37], site-controlled strain-induced excitation[8], in-situ cathodoluminescence (CL) lithography[9], and optical trapping[3,5]), introducing a powerful approach to improve our control on deterministic coupling and at the same time to facilitate scalable fabrication.

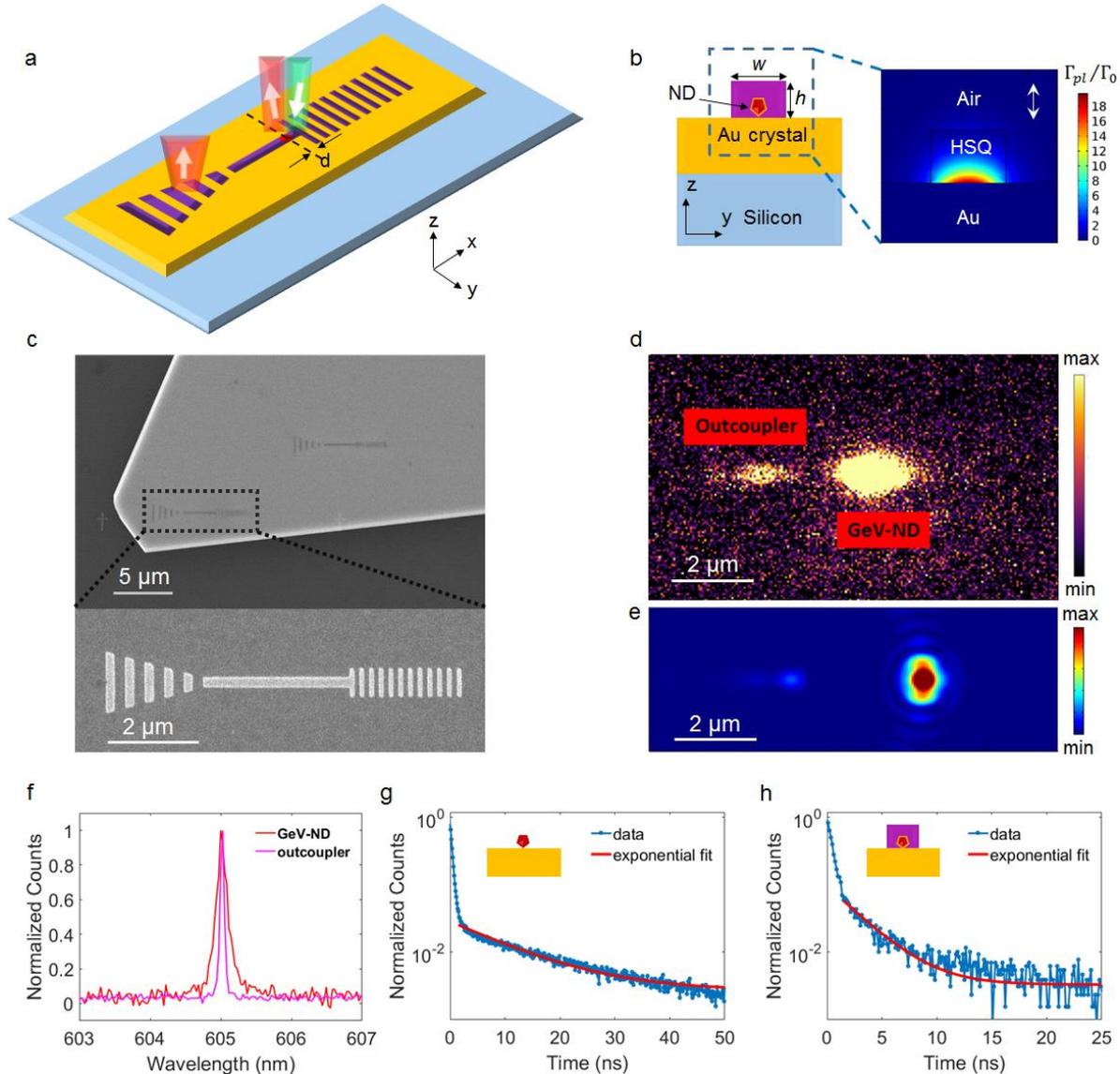

**Figure 3** Unidirectional excitation of a single GeV color center in a ND. (**a**, **b**) Schematic of the device layout and working principle. The dashed line in (**a**) shows *x* position of the embedded ND at the distance of the second constructive point of the RBG mirror, i.e., $d = 3\lambda_n/4$, where $\lambda_n$ is the wavelength of the SPP mode. Mode profile shown in (**b**) indicates distribution of Purcell enhancement ($\Gamma_{pl}/\Gamma_0$, plasmonic decay rate) for the GeV coupled system, while $w = 250$ nm and $h = 180$ nm. (**c**) SEM image of the fabricated device on gold crystal. The periodicity of the RBG is 240 nm. (**d**) Galvanometric mirror scan image of the coupled system. (**e**) Simulated result for an electric dipole oriented along *z*-axis to excite the propagating mode in the waveguide. The distance between the dipole and the RBG mirror was set as $d = 3\lambda_n/4$ for an efficient unidirectional excitation while the height of the dipole was set to 30 nm. (**f**) Spectra taken from the embedded GeV-ND (red) and from the outcoupled light at the end (purple). (**g-h**) Lifetime measurement data from uncoupled GeV (**g**, ND on gold crystal), coupled GeV (**h**, ND embedded in device). The data were analyzed with a single exponential function ($I_{tot} = A \cdot \exp(-t/\tau_A) + C$, where $\tau_A$ indicates the lifetime, and $A$ and $C$ are constants[16,30]), indicating a 5-fold lifetime shortening from 11.2 ns (ND on gold flake) to 2.3 ns (ND embedded in device). Considering an additional 2-fold reduction due to the metal layer[30], we estimated a 10-fold decay rate enhancement in total.

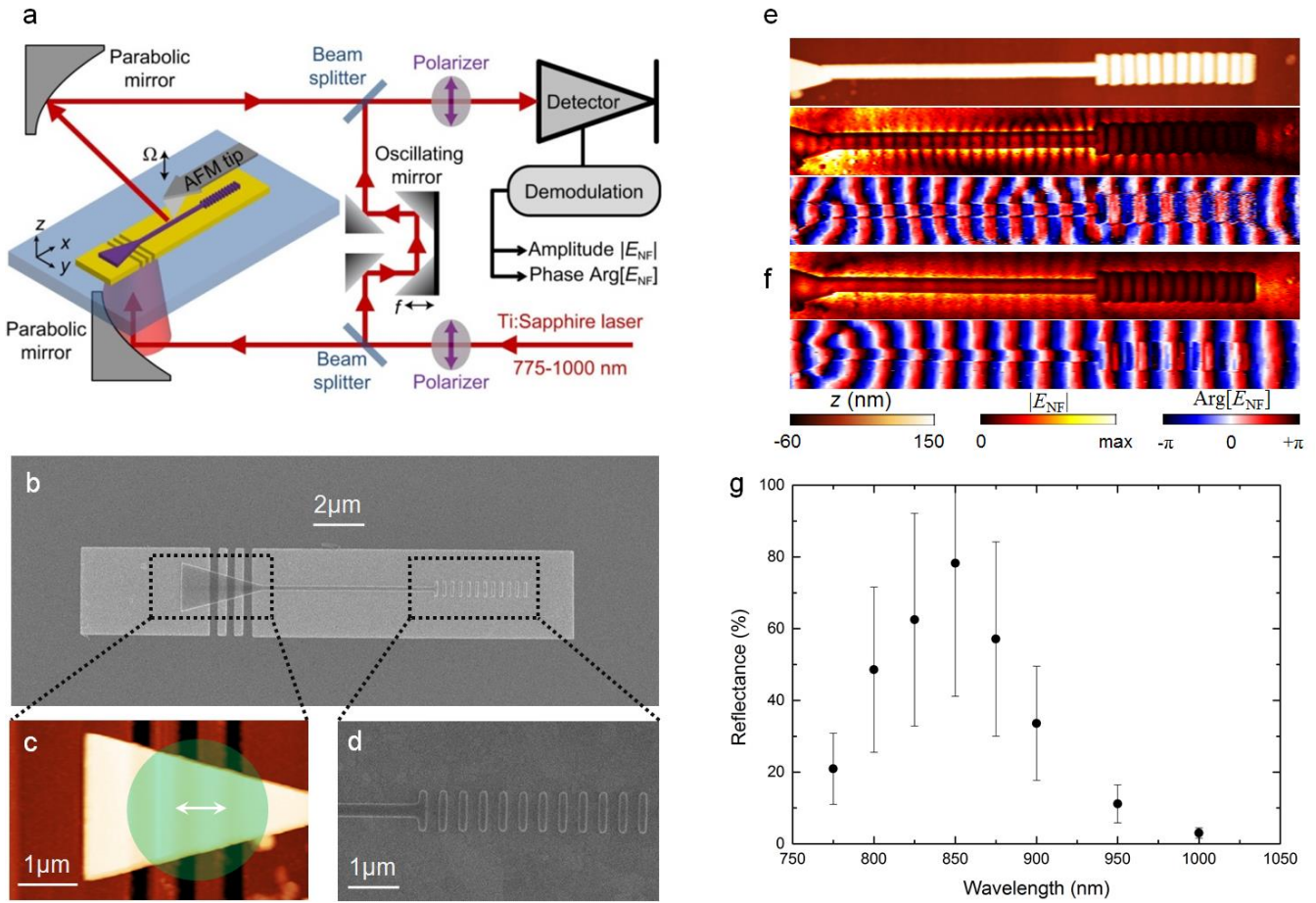

**Figure 4** Near-field investigation of the unidirectional SPP coupler (**a**) Schematic of the near-field optical setup. (**b**) SEM image of the fabricated device, namely dielectric nanoridge atop a patterned gold rectangular layer. (**c**) AFM image of the input grating, overlapped with dielectric funnel for excitation of dielectric-loaded SPP mode. Green circle and white arrow illustrate approximate position of incident illumination spot (not to scale) and its polarization, respectively. (**d**) Zoomed-in SEM image of RBG. (**e**, **f**) Topography z (top), near-field amplitude $|E_{NF}|$ (middle), and phase Arg $[E_{NF}]$ (bottom) of the unidirectional SPP coupler, recorded at λ = 850 nm (**e**, inside the TM bandgap) and λ = 1000 nm (**e**, outside the TM bandgap). (**g**) Reflectance of RBG, evaluated from SNOM maps.